\documentclass[12 pt]{article}
\usepackage{rotating}
\usepackage{subcaption}
\usepackage{bbm}
\usepackage{graphicx, caption,amsmath, enumitem, setspace,graphics, comment, amstext,lscape, hyperref}
\usepackage[margin=1.25in]{geometry}
\usepackage[usenames, dvipsnames]{color}
\usepackage[longnamesfirst]{natbib}
\usepackage{rotating}
\bibpunct{(}{)}{;}{author-year}{}{,}
\usepackage[margin=9pt,font=small,labelfont=bf,labelsep=endash]{caption}
\usepackage{booktabs}
\usepackage{setspace}
\usepackage{footmisc}
\usepackage{amsfonts}
\usepackage{lipsum}

\title{Principled estimation of regression discontinuity designs}

\author{Jason Anastasopoulos \\
\url{ljanastas@uga.edu} \thanks{I am very grateful to George Krause, Mariliz Kastberg-Leonard, Kosuke Imai, Chris Winship,  Gary King, Max Gopelrud,  Molly Offer-Westort,  Erin Hartman, Marc Ratkovic,  Kevin Esterling, Luke Miratrix , Richard Nielsen and Rocio Titunik for their helpful comments and assistance. This is a  draft, please do not cite without permission.}}

\date{First Version: August 30th 2018\footnote{Prepared for the annual \textit{American Political Science Association Conference} in Boston, MA} \\  Current draft: \today}

\begin{document}
\maketitle

\begin{abstract}
 Regression discontinuity designs are frequently used to estimate the causal effect of election outcomes and policy interventions.  In these contexts, treatment effects  are typically estimated with covariates included to improve efficiency. While including covariates improves precision asymptotically, in practice, treatment effects are estimated with a small number of observations, resulting in considerable fluctuations in treatment effect magnitude and precision depending upon the covariates chosen. This practice thus incentivizes researchers to select covariates which maximize treatment effect statistical significance rather than precision.
Here, I propose a principled approach for estimating RDDs which provides a means of improving precision with covariates while minimizing adverse incentives.
This is accomplished by integrating the adaptive LASSO, a machine learning method, into RDD estimation using an \texttt{R} package developed for this purpose, \texttt{adaptiveRDD}. 
Using simulations, I show that this method significantly improves treatment effect precision, particularly when estimating treatment effects with fewer than 200 observations.
  
\end{abstract}


\vspace{.5cm}
\textbf{Word count:} 6,348.  

\doublespacing
\newpage

 Regression discontinuity designs (RDDs) are often used in political science research to estimate the causal effect of close election outcomes  (see eg. ~\cite{caughey2011elections,erikson2017much,green2009testing,imai2011introduction,skovron2015practical}). 
The premise of the RDD is conceptually simple and intuitive.
Around a narrow interval of a threshold for a variable that assigns a treatment (running variable), treatments can plausibly be considered to be ``as-if'' randomly assigned.
While bandwidth selection, kernel choice and estimation strategy for RDDs are well understood, work on theoretical considerations regarding the common practice of including covariates to adjust local average treatment effect (LATE) estimates is relatively recent~\citep{frolich2007regression,calonico2019regression}. ~\cite{calonico2019regression} in particular provide strong theoretical grounds for continuing the practice of estimating RDDs with pre--treatment covariates. 

While including covariates improves treatment effect precision asymptotically, in practice, treatment effects estimated with RDDs are often done with a small number of observations, resulting in considerable fluctuations in treatment effect magnitude and precision depending upon the covariates chosen. As a result, this practice creates incentives for researchers to select covariates in a manner which maximizes the statistical significance, rather than precision,  of the treatment effect (``p-hacking''). Here, I propose a principled approach for estimating RDDs with covariates which provides a means of maximizing precision while minimizing adverse researcher incentives, particularly in small N contexts, by integrating the adaptive LASSO, a regularization method used in machine learning, into regression discontinuity design estimation. This approach is flexible and allows researchers to combine substantive knowledge with an automated covariate selection algorithm that is tailored to RDDs and used here for its  model selection (oracle) properties~\citep{zou2006adaptive}.\footnote{As I describe in more detail below, this contrasts with the more ``traditional'' version of the LASSO developed by~\cite{tibshirani1996regression} which is concerned primarily for MSE reduction at the expense of consistent model selection and specification.}. 

The remainder of this paper is as follows. Section 1 provides a brief introduction to LATE estimation for sharp RDDs with local linear regression (LLR), the focus of this paper;  Section 2 introduces the adaptive LASSO and accompanying implementation algorithm along with relevant theoretical derivations;  Section 3 provides an applied example of enhanced LATE estimation using a close election RDD study of the effect of holding political office on profit margins in Russian firms published in the \textit{American Political Science Review} by~\citep{szakonyi2018businesspeople}.
Section 4 provides empirical evidence of the bias reduction and efficiency gains of this method using a series of simulated close election RDDs with covariates and finally, Section 5 concludes with a discussion of future research in this area.  

\section{Covariate adjusted LATE in regression discontinuity designs}

Regression discontinuity designs are a framework for the causal estimation of local average  treatment effects with observational data.
This is accomplished using a running variable $F_i; i = 1,\cdots,n$ which assigns treatment $T_i$ on the basis of some threshold value $f$ such that if $F_i > f$, a unit (individual, geographic unit etc) is assigned to treatment $T_i = 1$ and is not assigned to treatment otherwise $T_i = 0$.
Assuming continuity of the forcing variable, the sharp RDD leverages this mechanism by allowing for the causal estimation of LATE around a narrow window of the threshold $f - \epsilon < f< f+\epsilon$ by making the assumption that, in the limit of this window, units are as ``as if'' randomly assigned to a treatment~\citep{hahn2001identification}. In line with other work on the RDD, this paper is concerned primarily with the sharp RDD, the most commonly used design in the political science and public policy literatures~\citep{calonico2016regression}.

Under the potential outcomes framework~\citep{rubin2005causal}, define $Y_i$ as the observed outcome for \textit{i}, $Y_{i}(1)$ as the outcome, had unit $i$ received treatment, and $Y_{i}(0)$ as the outcome had unit $i$  not received treatment, RDDs allow us to estimate the local average treatment effects (LATE) at the threshold $F_i = f$:\footnote{For the purpose of illustration, we assume that $f= 0$.} 

\begin{equation}\label{eq:tau}
 LATE = \displaystyle \tau = \lim_{F_{i} \downarrow 0} E[Y(1)_{i} | F_{i}  = f + \epsilon] - \lim_{F_{i}  \uparrow 0} E[Y(0)_{i} | F_{i} = f - \epsilon]
\end{equation}

Estimation of $\tau$ is typically accomplished through a local linear regression (LLR) in a neighborhood of the cutpoint $F_{i} \in [c - h,c + h]$ which is  determined through optimal bandwidth selection procedures designed to minimize cross-validated MSE~\citep{imbens2012optimal}.

\begin{equation}\label{eq:rddest}
		\hat{Y}_{i} = \beta_{0} + \hat{\tau} T_{i} + \gamma F_{i} + f(T_{i}, F_{i})
\end{equation}

In Equation~\ref{eq:rddest}, $\hat{\tau}$ is the estimated local average treatment effect, $T_{i}$ is a binary treatment indicator function which equals 1 when $F_{i} > 0$ and 
$f(T_{i}, F_{i})$ is a function of the forcing variable which often takes the form of a non-parametric kernel or $p^{th}$ order polynomial.
A common LLR model estimated in the literature is the model shown in Equation~\ref{eq:rddcov}~\citep{calonico2019regression}:

\begin{equation}\label{eq:rddcov}
		\hat{Y}_{i} = \beta_{0} + \hat{\tau} T_{i} + \gamma F_{i} +  \delta F_{i}\cdot T_{i} + X\beta
\end{equation}

In Equation~\ref{eq:rddcov},  a set of covariates \textit{X} are added to increase the precision of LATE.  ~\cite{calonico2019regression} derive the covariate adjusted estimator of $\hat{\tau}$ and demonstrate that pre--treatment covariate adjustment typically leads to more efficient estimates of $\hat{\tau}$ but, as mentioned above, there is little guidance regarding \textit{which} pre--treatment covariates should be included to maximize the efficiency of LATE.
Table~\ref{tab:1} which lists the types of covariates chosen for similar  close-election RDD designs highlights this problem.
This is particularly problematic in small N estimation contexts and when covariates are correlated with the running variable, cases in which covariate selection can have a much greater impact on LATE efficiency and point estimates. 
In these cases, which are very common in the political science literature, 

\begin{sidewaystable}
\centering \footnotesize
\caption{Covariate types chosen for RDD estimation in the ``top 3'' political science journals. ``Lowest N'' is the smallest number of observations used to estimate a RDD treatment effect in each paper.}
\label{tab:1}
\begin{tabular}{p{0.2\linewidth}p{0.2\linewidth}p{0.1\linewidth}p{0.2\linewidth}p{0.2\linewidth}p{0.05\linewidth}}
\hline \\
Journal (Year),Author(s) & Title &  DV & Forcing  & Covariate Type & Lowest N \\ \hline
\vspace{.1cm} \\
APSR (2018), Szakonyi &	``Businesspeople in elected office: Identifying private benefits from firm-level returns'' & Revenue, profits. & Vote margin. & Sector, region, year fixed effects, candidate level covariates. & 136 \\
APSR (2015), Hall & ``What happens when extremists win primaries?'' &	Party victory. &	Vote share margin. & Congress fixed effects. & 35 \\
JOP (2014), Boas, Hidalgo, and Richardson & ``The spoils of victory: campaign donations and government contracts in Brazil'' & Contracts. & Vote  margin. & Firm fixed effects. & 	45 \\
APSR (2014), Ferwerda and Miller & ``Political devolution and resistance to foreign rule: A natural experiment'' & Attacks. & Commune distance from demarcation line. & Mean elevation, train station distance, communications available, farmed area, ruggedness of the landscape, population. &	15 \\
AJPS (2011), Boas and Hidalgo &	``Controlling the airwaves: Incumbency advantage and community radio in Brazil'' & Radio station coverage. & Vote margin. & Population.	& 33 \\
APSR (2009),  Eggers and Hainmueller & ``MPs for Sale? Returns to Office in Postwar British Politics'' &	Logged wealth.  & Vote share margin. &	Candidate traits. & 165 \\ 
\vspace{.1cm} \\
\hline\hline

\end{tabular}
\end{sidewaystable}

As a solution to a similar problem in the context of randomized experiments~\cite{bloniarz2016lasso} propose selecting covariates using a shrinkage and variable selection  method known as the LASSO, a practice which I modify and extend to LATE estimation in the regression discontinuity design here by employing the adaptive LASSO, a version of the LASSO which has demonstrated oracle (correct model selection) properties~\citep{zou2006adaptive}. 

Covariate selection  using the adaptive LASSO has a number of benefits. 
First, given any initial set of covariates chosen by the researcher, subsequent covariate selection using this method can improve optimal bandwidth choice via model MSE minimization independent of the bandwidth estimation algorithm; second, this method can maximize LATE efficiency and; third, the method constrains the extent which a treatment effect estimate can be ``p-hacked'' through the practice of adding covariates. Each of these properties are demonstrated below.

\section{Regularization, machine learning and variable selection}

Regularization methods are tools used primarily for prediction problems and machine learning applications as a means of reducing the dimensionality of a feature space to avoid over fitting of a prediction model. 
In the context of linear models, ridge regression and lasso regression are the primary regularization methods used for linear prediction problems~\cite{tibshirani1996regression}.
Each method applies a term which penalizes each additional variable added to an OLS model in a different way. 
For instance, in all OLS problems our goal is to find coefficient estimates $\beta$ which minimize the squared error loss: 

$$
\mathbf{\hat{\beta}}_{OLS} = \arg\min_{\beta} \sum_{i = 1}^{N}(Y_{i} - X\beta)^{2}
$$
OLS under mild assumptions is guaranteed by Gauss-Markov to be the best linear unbiased estimator  (BLUE) of the coefficient values. 
However, if our ultimate goal is \textit{prediction} using a linear model, as is typically the case in the machine learning context,  the bias--variance trade-off allows us to exchange unbiasedness of coefficient estimates for a model that makes better out-of-sample predictions (lower MSE)~\citep{tibshirani2015statistical}.
This was first demonstrated by statistician and mathematician Charles Stein in 1956 and improved upon by statistician Williard James and Stein in 1961  and came to be known as James-Stein \textit{shrinkage} estimation of  linear models~\citep{stein1956,james1961proc}.

\subsection{Shrinkage and Regularization}

As its name suggests, shrinkage estimation is a means of optimizing the predictive abilities of linear models through shrinking coefficient estimates toward zero. One of the first shrinkage methods developed for linear models was ridge regression which added a $L_2$ penalty to the OLS minimization problem~\citep{tihonov1963solution}:

\begin{equation}\label{eq:ridge}
\mathbf{\hat{\beta}}_{Ridge} = \arg\min_{\beta} \underbrace{\sum_{i = 1}^{N} (Y_{i} - X\beta)^{2}}_{OLS}- \underbrace{\lambda \sum_{j = 1}^{p} \beta_{j}^{2}}_{Ridge~Penalty~(L_{2})}
\end{equation}

\noindent In Equation~\ref{eq:ridge} above, the original OLS loss function is estimated with a penalty term which penalizes the inclusion of additional variables and is determined by the tuning parameter $\lambda$ which is estimated using cross-validation~\citep{tibshirani1996regression}. 

This ridge regression estimator ends up introducing biased (shrunken) coefficient estimates, but through the introduction of this bias, minimizes MSE and improves ability of the model to make better predictions in out of sample data. 
Unfortunately, ridge regression cannot be used as a variable selection tool because it will never shrink coefficients to zero~\citep{tibshirani2015statistical}. However, the LASSO, an acronym for ``least absolute shrinkage and selection operator,'' which slightly modifies the penalty term above to an $L_{1}$ norm allows the model to serve as both a shrinkage and selection method:

\begin{equation}\label{eq:lasso}
\mathbf{\hat{\beta}}_{lasso} = \arg\min_{\beta} \underbrace{\sum_{i = 1}^{N}(Y_{i} - X\beta)^{2}}_{OLS~Loss} - \underbrace{\lambda \sum_{j = 1}^{p}  |\beta_{j}|}_{Lasso~Penalty~(L_{1})}
\end{equation}

The nature of the constrained optimization problem presented by the objective function in Equation~\ref{eq:lasso}, some coefficients will be shrunk toward zero, thus allowing for the LASSO to be model selection and shrinkage tool~\citep{tibshirani2015statistical}.
Additional versions of the LASSO which involved tweaks to the penalty for specific  high dimensional problems include the elastic net, which combines ridge and LASSO penalties, and the ``group lasso'', which is used to select out large groups of covariates~\citep{meier2008group,simon2013sparse}.

\subsection{Variable selection and oracle properties of the adaptive lasso}

Most variations of the LASSO applicable to high dimensional ($p>n$) data often do a good job of minimizing MSE, but fare poorly in simulations in which the ultimate goal is to retrieve the correct subset of covariates from a relatively large pool~\citep{zou2006adaptive}. 
As such, the usefulness of the standard LASSO for LATE adjustment in RDDs, which do not typically involve high dimensional problems with covariates, is somewhat questionable. 
Fortunately, the \textit{adaptive LASSO} introduced by~\cite{zou2006adaptive} was developed with the goal of maximizing ``correct'' variable selection for both low and high-dimensional estimation problems, making it an ideal candidate for selecting covariates in RDDs and other causal inference contexts in which covariate adjustment is appropriate. 
As with other flavors of the LASSO the adaptive LASSO requires adjustment of the penalty term:

\begin{equation}\label{eq:adaptive}
\mathbf{\hat{\beta}}_{adaptive} = \arg\min_{\beta} \underbrace{\sum_{i = 1}^{N}(Y_{i} - X\beta)^{2}}_{OLS~Loss} - \underbrace{\lambda \sum_{j = 1}^{p}  \omega_{j}|\beta_{j}|}_{Adaptive~Penalty~(L_{1})}
\end{equation}

In Equation~\ref{eq:adaptive}, the inclusion of a set of weights $\mathbf{\omega}$, differentiates the adaptive LASSO from other LASSO varieties. For the adaptive LASSO, weights are chosen from the OLS estimates of the coefficients such that:

\begin{equation}\label{eq:weights}
    \displaystyle \omega_{j}  = \frac{1}{|\beta_j|^{\gamma}}
\end{equation}

\noindent where the $\beta_j$ are the coefficients estimated from an OLS model and $\gamma > 0$ is a tuning parameter\footnote{In~\cite{zou2006adaptive}, $\gamma$ was tuned using cross-validation and set to 0.5,1 and 2. In his simulations, the best results were achieved with $\gamma = 2$ followed by  $\gamma$ selected by cross validation. The tuning parameter $\lambda$  is estimated in the ordinary way via \textit{k}-fold cross-validation. In most software packages \textit{k} is set to 10 but this should be adjusted depending upon sample size. In the \textbf{R} software developed for this application, the default value of $\gamma$ is 2 but the user can choose to use change $\gamma$ using cross-validation or to another value of their choosing}:

$$
Y_{i} = \beta_{0} + \beta_{1}X_{1} + \cdots + \beta_{j}X_{j}
$$

What makes the adaptive LASSO appealing for causal inference, in general,  is that with the appropriate value of $\lambda$ estimated from the data, the adaptive lasso exhibits oracle properties: it tends to consistently select a correct subset of variables out of a larger set and has asymptotic guarantees of unbiasedness and normality~\citep{zou2006adaptive}. This is especially useful when the lasso is used as a variable selection, rather than shrinkage tool, which will be true more often in the context of covariate adjustments of LATE in RDDs and other causal inference contexts more generally. 

Indeed, as with ridge regression and other varieties of lasso, however, raw parameter estimates ($\hat{\beta}_{adaptive})$  can be biased in finite samples, which may appear to limit the utility of this method for causal inference more generally. 
Fortunately, however, as ~\cite{bloniarz2016lasso}, ~\cite{wager2016high} and others point out, estimation through a two-step procedure in which the lasso is used as a model selection tool and final parameter values are estimated using OLS allows us to obtain BLUE coefficient estimates with appropriate standard errors in an easily interpretable model.

Accordingly, this  is the approach that I employ here that is discussed in more detail below.
Furthermore, here, as in~\cite{bloniarz2016lasso}, we argue that adaptive lasso covariate adjustment of LATE can improve the precision of estimates and also function as a means of ``principled'' model selection that can avoid some of the pitfalls of model manipulation to recover statistically significant treatment effects (ie ``p-hacking'') for RDDs.
Based on a series of simulations and on the basis of the theoretical results discussed here and previously in~\citep{bloniarz2016lasso}, we recommend a four--step  process for RDD treatment effect estimation when covariates are included. This process is outlined in Table~\ref{tab:concept} and described in more detail below.

\subsection{Principled RDD estimation algorithm}
\begin{table}[ht!]
\centering 
\footnotesize
\begin{tabular}{lll}
\hline \\
\textbf{Step 1} & Researcher pre-treatment  & Covariates are selected by the researcher \\ 
		 & covariate selection		& on the basis of substantive concerns. \\
                 & & and data limitations. \\
                 & & \\
\textbf{Step 2} & Adaptive lasso regularization & The model from Step 1 is estimated using an adaptive lasso \\ 						&  & as described below.  \\
  & & \\
\textbf{Step 3} & Covariate adjustment & Covariates and higher-order terms whose coefficients are  \\
			& & shrunk to 0 are excluded from the final model. \\
            & & \\
            & & The adaptive lasso is tailored in this case \\
            && such that \textbf{the treatment effect, forcing variable} \\ 
            & & \textbf{and variables included in the kernel chosen are}  \\ 
            & & \textbf{NOT penalized. } \\
                        & & \\
\textbf{Step 4} &  CCT robust estimation  & The modified model from Step 3 is estimated \\ 
			   & of final model	 & via the CCT robust procedure \\
                & & \citep{calonico2014robust}.  \\ \hline \hline
\end{tabular}
\caption{Overview of the principled RDD estimation algorithm with the adaptive LASSO.}
\label{tab:concept}
\end{table}

Briefly, the four steps involve researcher model selection based on substantive or theoretically motivated concerns, the application of a adaptive lasso regularization with tuning parameter cross validation; variable selection based on the results of adaptive lasso estimation in the previous step and finally CCT robust estimation of the model selected from Step 3. Each of these steps along with treatment effect estimates produced by this method in the context of RDDs with local linear regression and covariates are derived below.

\section{Adaptive lasso estimation of LATE for RDDs}

\subsection{Step 1: Researcher Pre-Treatment Covariate Selection }

The purpose of including pre-treatment covariates in RDD estimation, as in randomized experiments, is to increase the precision of treatment effect estimates~\citep{bloniarz2016lasso,calonico2018regression}. This increase in precision can be the result of improved bandwidth selection, reduced model variance or a combination of the two. Some questions that researchers may struggle with, however, are: (1) \textit{which} pre-treatment covariates to include in the model and; (2) whether pre-treatment covariates should be included before or after optimal bandwidth selection. 

This is a thorny issue because all of these decisions can have significant downstream consequences for LATE estimation and efficiency, particularly when covariates included are highly correlated with the forcing variable and in small N local linear regression contexts which tend to be common in RDD estimation within the political science literature. As a result, temptations to manipulate covariate selection to maximize the statistical significance of LATE estimates is high, particularly in cases where LATE estimated without covariates are marginally significant (i.e. $0.05 < p < 0.10$).

While the automated model selection algorithm proposed here in Table~\ref{tab:concept} cannot eliminate ``p-hacking'', it is a procedure that can at the very least attenuate the ability of researchers to engage in this practice while simultaneously providing LATE estimates when covariates are introduced than researcher model selection alone. 
That being said,  initial decisions regarding which pre-treatment covariates to include should \textbf{always} be made on the basis of expert judgment/the researcher's expectation of which are the most relevant to the problem at hand. 
Since RDDs in the political science literature are typically conducted with close election vote share as the forcing variable $F_i$ and the treatment of interest is an election win $T_i$ where $T_i = 1$ if $F_{i} > c$ and $T_i = 0$ otherwise we focus on this type of RDD to illustrate the method. 

\begin{equation}\label{eq:origrdd}
			Y_{i} = \alpha + \tau T_{i} + \gamma F_{i} +  \delta (F_{i} \cdot T_{i}) + X\beta + \epsilon_{i}
\end{equation}
Equation~\ref{eq:origrdd} is a typical local linear regression model estimated to obtain the treatment effect estimate $\hat{\tau}$  where the observations $i$ are in some neighborhood, $a$  of the forcing variable $F_i$ around the cutpoint $c$, i.e. $i \in F_{a} \pm c$ and $X$ is a matrix of covariates. In these circumstances, the covariates included are often characteristics of the candidate (age, sex, etc) and characteristics of an electoral unit that they represent (pre-treatment demographics etc). ~\cite{szakonyi2018businesspeople}, for instance includes candidate controls such as age, gender, incumbency, ruling party membership, state ownership, foreign ownership, and logged total assets in the pre-election year in his estimates. As I mentioned above, selection of this initial set of covariates should \textit{always be dictated by a substantive understanding of the problem at hand.}

\subsection{Step 2: Adaptive lasso regularization}

Once the model in Equation~\ref{eq:origrdd} has been selected, a question that remains is whether this is the \textbf{best} possible model that can be fit which invariably raises the question of what ``best'' means in this context. Here we define ``best'' as a model in which a set of covariates $X^*$ are chosen out of the original set of covariates $X$ which minimizes the variance of  LATE, $Var(\hat{\tau})$. All things equal, it can be shown that minimizing $Var(\hat{\tau})$ can be accomplished by minimizing the mean squared error (MSE) of the local linear regression. 

Formally, if $X^{s}$ is a subset of covariates from $X$, we seek to choose an $X^{s} \subseteq X$ such that:

\begin{equation}\label{eq:var}
Var(\hat{\tau}| X^{s}) \leq Var(\hat{\tau}| X)
\end{equation}

\noindent Describing as the vector of coefficients that we estimate in the simple sharp RDD case as $\mathbf{\Theta} =  (\tau, \gamma, \delta, \beta)$ we seek:

\begin{equation}\label{eq:minset}
\displaystyle  \arg\min_{\mathbf{\Theta}} \sum_{i = 1}^{N}( Y_{i} - [\alpha + \tau T_{i} + \gamma F_{i} +  \delta (F_{i} \cdot T_{i}) + X^{s}\beta])^{2}
\end{equation}

While many methods exist for choosing $X^{s}$, LASSO regularization is well suited to the estimation of linear models and has been found to outperform other automated variable selection methods~\citep{tibshirani2015statistical}. Also, since we are primarily concerned with optimal model selection in a relatively small-N context, the adaptive LASSO is a natural choice since it is the only lasso variety which possesses the oracle property, as mentioned above. This is important because it guarantees that it will be consistent in both estimation of $\tau$ and in variable selection. Formally this implies asymptotic unbiasedness of $\hat{\tau}$ in the ordinary sense:

$$
\sqrt{n}(\hat{\tau}  - \tau) \rightarrow N(0, \mathbb{I}^{-1}(\tau))
$$
\noindent while simultaneously identifying the correct set of non-zero coefficients.  These properties ensure that adaptive lasso estimates of $\tau$ are asymptotically \textit{at least as good, in terms of efficiency and bias, as LLR without adaptive lasso variable selection}.

Learning about which covariates to exclude in RDDs, however,  requires modifying the adaptive lasso to the RDD context. In particular, we do not want to penalize the treatment effect, forcing variable or kernel, but do want to penalize any additional covariates. 
This can be accomplished by simply estimating a modified version of the adaptive lasso in which the weights for these coefficients are set to 0 while the weights of the added covariates are identical to those of the adaptive lasso.
The full initial model to be estimated is thus:

\begin{equation}\label{eq:minset}
\displaystyle  \arg\min_{\mathbf{\Theta}} \sum_{i = 1}^{N} \left [ Y_{i} - (\alpha + \tau T_{i} + \gamma F_{i} +  \delta (F_{i} \cdot T_{i}) + X\beta)\right]^{2}  +  \lambda\left[ \sum_{j  = 3}^{p} \omega_{j} |\beta_{j} | \right]
\end{equation}

Where $\omega_{j} = 1/|\beta_{j} |^{\gamma}$ are obtained through OLS estimation of $\beta_j$ and $\gamma$ is determined through cross-validation as described above. Again, the tuning parameter $\lambda$ is estimated with k-fold cross validation.

\subsection{Step 3: Automated Model Selection}

Once parameters from the adaptive lasso model in Equation~\ref{eq:minset}  are estimated using the optimal penalty value $\lambda^{RDD}$ and optimal weights, those covariates which are shrunk to zero are excluded from the model prior to calculating the optimal bandwidth. The resulting model used to estimate  optimal bandwidth and subsequently, robust treatment effects, is will thus be:

\begin{equation}\label{eq:minset}
\displaystyle  \mathbb{E}(Y_{i} | T_{i},F_{i}, X^{s})  = \alpha + \tau T_{i} + \gamma F_{i} +  \delta (F_{i} \cdot T_{i}) + X^{s}\beta
\end{equation}

\noindent where $X^{s} \subseteq X$ is the truncated set of covariates selected out by the adaptive lasso described above. 
Since optimal bandwidth selection algorithms such as  Imbens--Kalyanaraman use cross-validated MSE as criteria for selecting the ``best'' possible bandwidth,  MSE for bandwidth values estimated using covariates pre-processed by the adaptive lasso method described should be equal to or less than model MSE for bandwidth values estimated using the full model from Step 1. 

As I demonstrate below, this method can be incorporated into RDD estimation with covariates \textit{before} bandwidth selection, which will alter the optimal bandwidth chosen, or \textit{after} bandwidth selection if the bandwidth is set to a predetermined value (eg. 1\%, 5\% etc for close election RDDs).

\subsection{Step 4: Regularized CCT Robust Estimation }

Steps 1-3 involve selecting an optimal LLR conditional expectation function (CEF),  $\mathbb{E}(Y_{i} | T_{i},F_{i}, X^{s})$,  and estimating an optimal bandwidth $h^{*}_{o}$ based on the CEF. Once this has been accomplished, final treatment effect estimates are produced using  CCT robust estimation~\citep{calonico2018regression}.

\section{Empirical Illustration: Do Firms Profit from Having Elected Board Members?}

Knowledge of whether politicians benefit financially from holding office-holding is essential is essential for ensuring the legitimacy of democratic institutions.  Earlier work using RDDs to estimate the returns to office found large lifetime earnings effects by barely (initially) elected members of the British Parilament~\citep{eggers2009mps}. Subsequent work in different national contexts has found similar results as well (see eg~\cite{fisman2014private} (India), \cite{truex2014returns} (China), etc).~\cite{szakonyi2018businesspeople} adds to this literature by using a close election RDD to explore whether office-holding affects the profits of firms whose board members held political office in Russia. 
Using a close election RDD, ~\cite{szakonyi2018businesspeople} finds that office holding positively affects both firm profitability and firm revenue. In the empirical illustration below, I replicate~\cite{szakonyi2018businesspeople}'s RDD with and without the principled estimation method discussed here.

In the following analysis, I replicate the local linear regression in Table 2 of~\cite{szakonyi2018businesspeople}. In this table, the author uses a close election RDD to estimate the causal effect of holding political office on firm profitability with and without covariates using a 5\% bandwidth as well as the Imbens-Kalyanaraman optimal bandwidth estimated without covariates.
The general form of the local linear regression estimated is:

\begin{equation}\label{eq:szllr}
\text{Firm Profits} = \alpha + \hat{\tau} (\text{District Win}) + \gamma \text{Margin} +  \delta (\text{District Win} \times \text{Margin}) + X\beta + \mathbf{Y}_{j} +  \mathbf{S}_{j} +\mathbf{R}_{j}
\end{equation}

In Equation~\ref{eq:szllr}, the outcome variable is firm profit margins and the treatment indicator is whether the businessperson won election in their district and the running variable is the vote margin. These analyses also include a set of covariates \textit{X} and year, sector and region fixed effects  (Y, S, R).  This regression is estimated around a threshold of the cutpoint $c \pm h^{full}$ where $c \pm h^{full}$ is determined through cross-validation. Define the original treatment effect of the full model (i.e. the model entered in Step 1 above), $\hat{\tau}(h^{full})$.

After selecting covariates via the adaptive LASSO through Steps 2-4, we are left with the model:

\begin{equation}\label{eq:regllr}
\text{Firm Profits} = \alpha + \check{\tau} (\text{District Win}) + \gamma \text{Margin} +  \delta (\text{District Win} \times \text{Margin}) + X^{s}\beta^{s} + \mathbf{Y}_{j} +  \mathbf{S}_{j} +\mathbf{R}_{j}
\end{equation}

Note that the primary difference between the two equations above is the new set of covariates  $ X^{s}\beta^{s}$ which satisfies the condition $rank(X^{s}) \leq rank(X) $ through the removal of covariates and a new optimal bandwidth $h^{optim}$ as a result of the addition of new covariates. While coverage properties of this new estimator is less clear theoretically, results from CCT and others suggest that the regularization adjusted estimator will have superior coverage properties as well under a variety of circumstances. This is confirmed in a series of simulations below.

\begin{table}[ht!]
\centering \footnotesize
\def\arraystretch{1.15}%
\begin{tabular}{rlllll}
  \hline
 						&Original &  Adaptive &    Original &    Adaptive   & Adaptive   \\
                         &(APSR)  &                      &    5\%  (APSR)      &  5\% & 	CCT Robust \\ \hline
                         & & & & & \\
  District Win & 0.146$^{***}$ & 0.102$^{*}$ & 0.198$^{**}$ & 0.097$^{**}$  &  0.140$^{***}$ \\ 
    					& (0.065)        &  (0.060)    &  (0.090)     & (0.038)       &  (0.052)			\\
                        & & & & & \\
   Bandwidth & 0.113           & 0.120        & 0.050        & 0.050     	 & 0.120		\\
   Covariates Dropped & * & 4 & * & 2  & 4 \\
     & & & & & \\
   Firm and Cand & Full & Select   & Full & Select		& Select		\\
   Covariates        &     &      &  		&		&					\\
   Region,Sector & Full & Full & No & No	& No					\\
   Year FE             &      & 		& & &													\\
   Observations  & 481 & 520 & 201 & 201 &  520     \\ \hline\hline 
\end{tabular}
\caption{\textbf{Replication of Political Connections and Firm Profitability Analysis in~\cite{szakonyi2018businesspeople} with Adaptive LASSO Adjusted Treatment Effects. } }
\label{tab:lassotab}
\end{table}

Table~\ref{tab:lassotab} contains original and regularization adjusted treatment effects and standard errors. One thing of note is that the standard errors of all adaptive lasso treatment effects are smaller than those of the original covariate adjusted treatment effects published. As simulations below demonstrate, this is due to the oracle property enjoyed by the adaptive Lasso, which has been demonstrated produce ``correct'' model specification under a wide variety of conditions.

\section{Simulations}

To compare the performance of the adaptive lasso method with conventional RDD estimates, I performed a series of simulations in which the number of observations were varied and the performance of each model was assessed in terms of treatment effect bias and percent coverage. The results of these simulations suggest that benefits gained for RDD estimation with the adaptive LASSO is greatest for smaller datasets ($<200$) but adaptive LASSO estimation outperforms conventional methods in simulations with larger datasets as well. 

To explore the bias and coverage properties of the adaptive LASSO method in a realistic applied scenario, I generate a series of simulated datasets using  mean and variance parameter estimates from real data. 
To accomplish this, I generate a series of simulated datasets using  the correlation matrix of the covariates and vote margin used by ~\cite{szakonyi2018businesspeople} to construct 2,000 simulated data sets which have the same covariance structure and mean of the original dataset and set the true treatment effect $\tau_{RDD}$ to 0.30.

Define $\mathbf{\Xi}$ as a matrix which contains the set of covariates plus the vote margin used in~\cite{szakonyi2018businesspeople} discussed above. Furthermore, assume that the data generating process of $\mathbf{\Xi}$ is that of a multivariate normal distribution defined by some mean parameters $\mathbf{\mu} = (\mu_{1}, \mu_{2},\cdots,\mu_{p})$ and a covariance matrix $\Sigma$. Thus:

$$
\mathbf{\Xi} \sim \mathcal{N}(\mathbf{\mu}, \mathbf{\Sigma})
$$
Using this data generating process along with empirically defined parameters $\mathbf{\mu}$ and covariance structure $\Sigma$, I generate $s = 1, \cdots, 2000$ simulated data sets  $\mathbf{\Xi}^{s}$ such that the d.g.p of each simulated dataset adheres to: 

$$
\mathbf{\Xi}^{s} \sim \mathcal{N}(\mathbf{\mu}, \mathbf{\Sigma})
$$
\noindent Through generating the data in this manner, we're insuring that each simulated dataset conforms to a realistic d.g.p in the context of a close-election RDD. For each simulation the outcome $Y$, and thus the true model, is thus defined by

$$
Y_{s} = 0.3(District~Win_{s}) + \gamma(Margin_{s}) + \delta(District~Win_{s} \times Margin_{s})  + \eta_{s}
$$
\noindent where the error term $\eta_{s} \sim N(0,1)$, the simulated vote share, $Margin_{s}$ is simulated as one of the variables within $\mathbf{\Xi}^{s}$ and $District~Win_{s} = \mathbb{I}(Margin_{s} > 0)$ is a simulated forcing variable based on $Margin_s$. The true treatment effect that we estimate using the conventional RDD approach and adaptive lasso approach with is $\tau_{RDD} = 0.3$. Reported coefficient values, bandwidths and standard errors are CCT robust estimates using the standard and adaptive approaches. 

The model estimated for each simulation is the full model including covariates:

\begin{equation}\label{eq:simulated}
\text{Y}_{s} = \alpha^{s} + \hat{\tau}^{s}_{RDD} (\text{District Win}_{s}) + \gamma^{s} \text{Margin}_{s} +  \delta^{s} (\text{District Win}_{s} \times \text{Margin}_{s}) + X^{s}\beta^{s}  + \epsilon^{s}
\end{equation}

\noindent For the simulations, the average bias of $\hat{\tau}^{s}_{RDD}$, $SE(\hat{\tau}^{s}_{RDD})$ and \% coverage of the confidence intervals were recorded for models in which the bandwidth was allowed to vary according to the adaptive lasso procedure outlined above or was fixed at a certain value with the adaptive lasso applied afterwards. 

\begin{figure}[ht!]
\centering
\includegraphics[width = .9\textwidth]{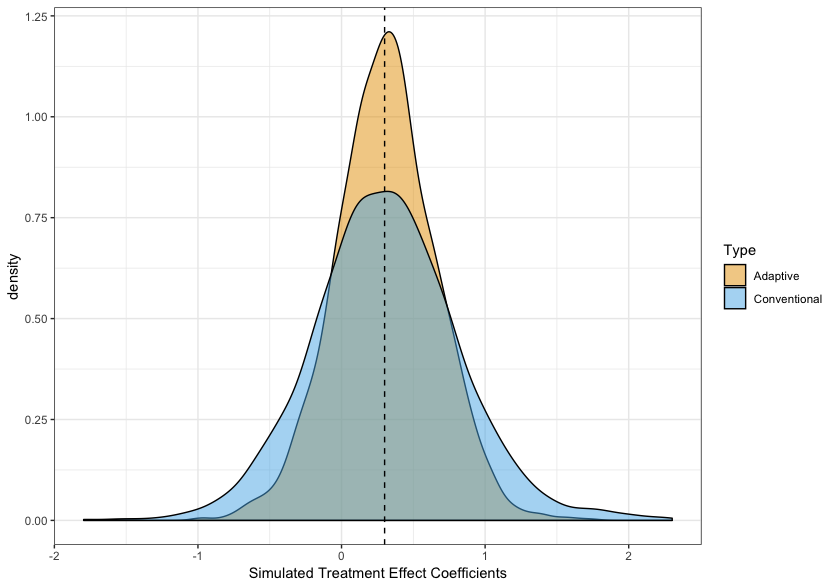}
\caption{Distribution of simulated treatment effects $\hat{\tau}^{s}_{RDD}$, for adaptive lasso adjusted treatment effects and conventional treatment effects across 2,000 simulated data sets with variable bandwidth select. The true $\tau_{RDD} = 0.30$ is denoted by the black dotted line.} 
\label{fig:simcoefbias}
\end{figure}

Figure~\ref{fig:simcoefbias} contains the distribution of simulated treatment effects estimated using conventional and adaptive lasso methods. Here we see that the adaptive lasso restricts the treatment effects estimated to a much narrower band around the true treatment effect.  

\clearpage
\begin{table}[ht]
\centering \footnotesize
\begin{tabular}{rllc}
\hline \\
&  & \textbf{Variable Bandwidth*} & \\ 
                          & & & \\
  						& \textit{Adaptive} & \textit{Conventional} &  \textit{Difference (Adaptive - Conventional)}   \\ 
$\tau_{RDD}$ Bias & 0.274 & 0.397 & -~0.123$^{***}$ \\ 
  \% Coverage & 0.944 & 0.699 & +~0.245$^{***} $ \\ 
  $\tau_{RDD}$ Estimate & 0.308 & 0.308 &  -  \\ 
  Bandwidth & 0.38 & 0.292 & +~0.088$^{***}$  \\ 
     				& & & \\ 
 &  &  \textbf{Fixed Bandwidth$^\mp$} & \\ 
    & & & \\
    & \textit{Adaptive} & \textit{Conventional} &  \textit{Difference (Adaptive - Conventional)}    \\ 
$\tau_{RDD}$ Bias   & 0.375 & 0.375 & -~0.001$^{~~~}$ \\
\% Coverage  & 0.931 & 0.796 & +~0.135$^{***}$ \\
 $\tau_{RDD}$ Estimate & 0.300 & 0.300 & -~0.001$^{~~~}$ \\
  Bandwidth & 0.200 & 0.200 &   -  \\
   \hline\hline
\end{tabular}
\caption{\textbf{Performance of Adaptive Lasso v. Conventional Treatment Effect Estimates in Simulations}$ ^{***}p < 0.01$, $ ^{**}p < 0.05$, $ ^{*}p < 0.10$ for t-test of mean difference with$H_{0}: \mu_{Adaptive} = \mu_{Conventional}$.  Average simulation results across 2,000 simulations comparing ``Adaptive'' vs. ``Conventional'' treatment effect bias and coverage results. Final bias and coverage results are both estimated using CCT robust point estimates and confidence intervals. *``Variable bandwidth'' results are produced through Imbens-Kalyanaraman optimal bandwidth selection based on models selected by the adaptive algorithm described above or the full model mentioned in this section. $\mp$ For fixed bandwidth simulations, bandwidth was set to 0.20. }
\label{tab:simeffects}
\end{table}

Table~\ref{tab:simeffects} contains estimates of the bias, \% coverage and other statistics from the simulation. The adaptive lasso here provides some very striking efficiency improvements which are reflected in the \% coverage estimates in both variable and fixed bandwidth selection procedures. In the variable bandwidth scenario, the adaptive lasso combined with CCT robust estimation produces confidence intervals on treatment effects that achieves an average of 94\% coverage versus 70\% coverage under conventional estimation while under the fixed bandwidth scenario, adaptive LASSO estimation achieved 93\% coverage compared to about 80\% coverage under conventional estimation. Each of these differences was statistically significant at the $p < 0.01$ level.

\subsubsection{Simulations by sample size}

\begin{figure}[ht!]
    \centering
    \includegraphics[width = .9\textwidth]{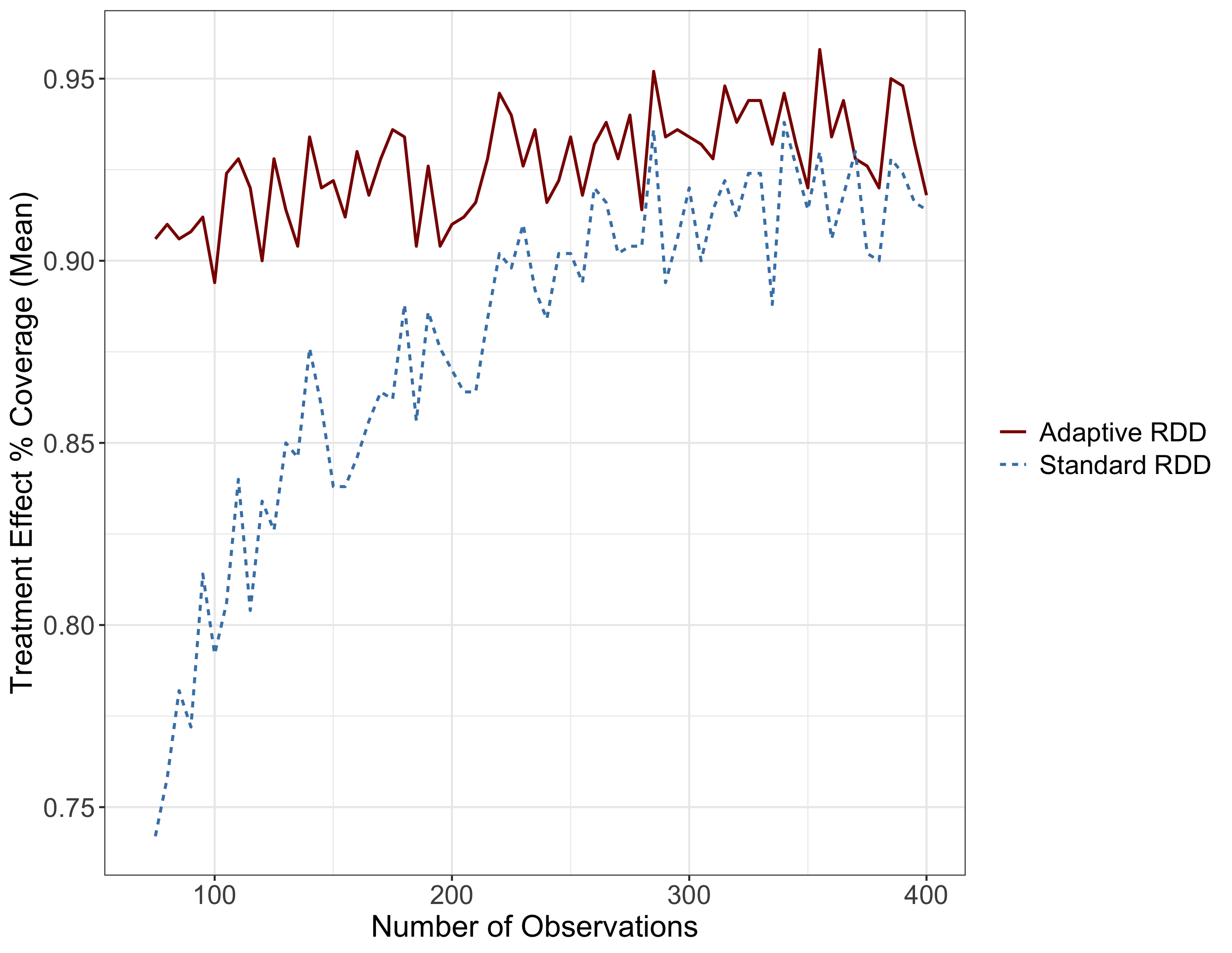}
    \caption{Mean \% Coverage by number of observations. For each N, 1,000 simulations were conducted and each point represents the mean \% coverage over each 1,000 simulations.}
    \label{fig:pctcoverageREALISTIC}
\end{figure}
\begin{figure}[ht!]
    \centering
    \includegraphics[width = .9\textwidth]{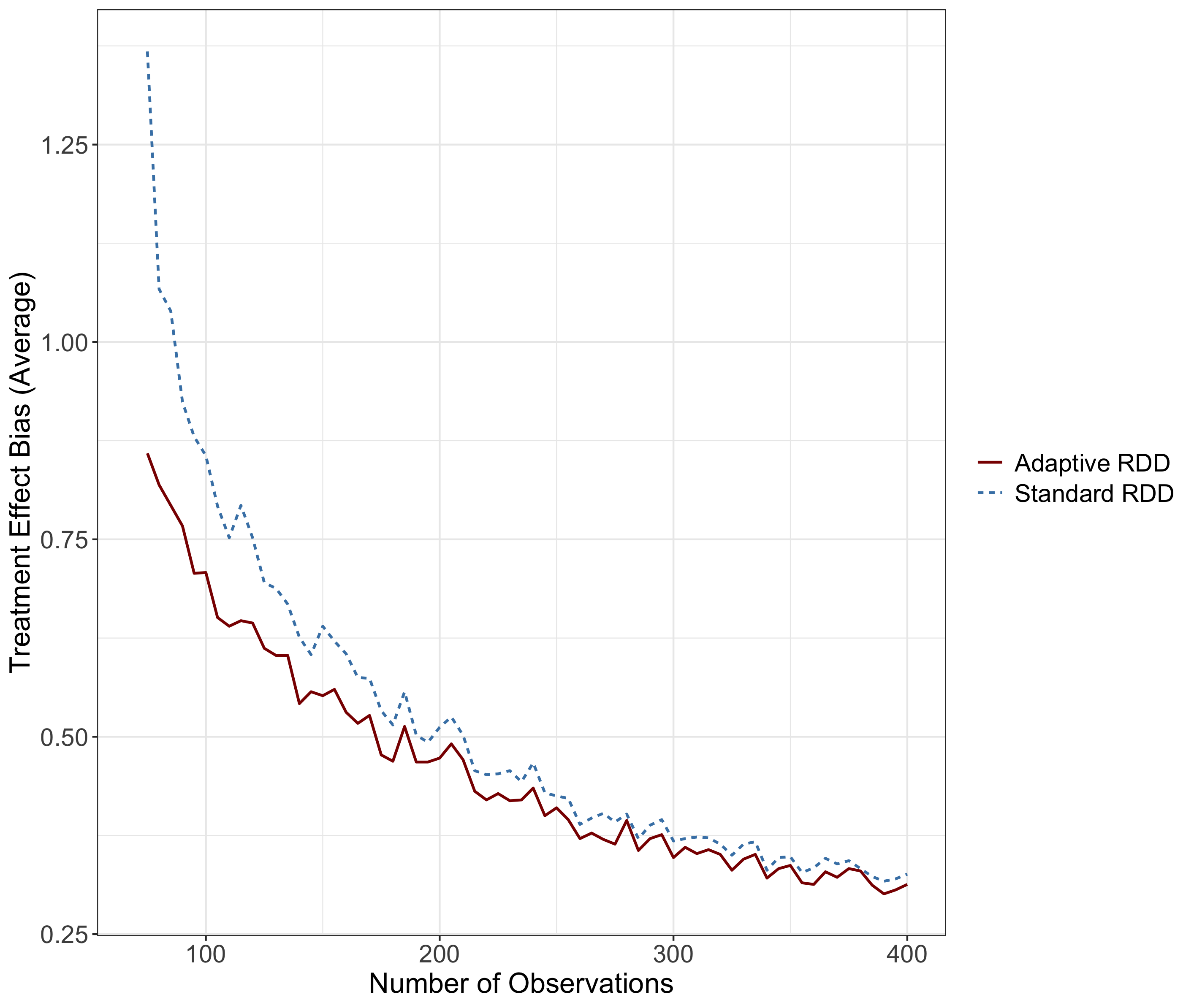}
    \caption{Mean bias by number of observations. For each N, 1000 simulations were conducted and each point represents the mean \% coverage over each 1,000 simulations.}
    \label{fig:biasREALISTIC}
\end{figure} 

To understand performance varies by sample size, I ran the same simulations described above 1000 times for between 70 to 400 by 5 and averaged treatment effect \%  coverage and bias for each number of observations around the cut point.  Figure~\ref{fig:pctcoverageREALISTIC} contains estimates of \% coverage by sample size and Figure~\ref{fig:biasREALISTIC} contains estimates of bias by sample size. 
These suggest that \% coverage is consistently better regardless of sample size but the improvement is most noticeable below 200 observations.  The same can be said of bias. 

\section{Discussion}

In this paper we have demonstrated that incorporation of the adaptive LASSO into RDD treatment effect estimation can  improve the efficiency of treatment effect estimates when covariates are included  and can also provide a principled framework of treatment effect adjustment for RDDs. Results of simulations included in these analyses suggest that this method is particularly useful when RDD treatment effects are estimated with fewer than 200 observations, which is when we strongly recommend that this method be used. As we emphasize above, however, this does not imply that substantive considerations in the estimation process should be abandoned and replaced by automated machine learning methods.
To the contrary, substantive considerations, as reflected in the algorithm that we developed above,  are and should always be at the forefront of model estimation whether in the context of RDDs or estimation strategies.

\clearpage

\bibliographystyle{apsr_fs} 
\bibliography{references.bib}

\begin{thebibliography}{xx}

\bibitem[Bloniarz et~al.(2016)]{bloniarz2016lasso}
Bloniarz, Adam, Hanzhong Liu, Cun-Hui Zhang, Jasjeet~S Sekhon, \harvardand\ Bin
  Yu. 2016.
``Lasso adjustments of treatment effect estimates in randomized experiments.''
  {\em Proceedings of the National Academy of Sciences} 113~(27): 7383--7390.

\bibitem[Calonico et~al.(2016)]{calonico2016regression}
Calonico, Sebastian, Matias~D Cattaneo, Max~H Farrell, \harvardand\ Roc{\i}o
  Titiunik. 2016.
``Regression discontinuity designs using covariates.'' {\em URL
  http://www-personal. umich. edu/\~{}
  cattaneo/papers/Calonico-Cattaneo-/Farrell-Titiunik\_2016\_wp. pdf}.

\bibitem[Calonico et~al.(2018)]{calonico2018regression}
Calonico, Sebastian, Matias~D Cattaneo, Max~H Farrell, \harvardand\ Rocio
  Titiunik. 2018.
``Regression discontinuity designs using covariates.'' {\em Review of Economics
  and Statistics}~(0).

\bibitem[Calonico et~al.(2019)]{calonico2019regression}
Calonico, Sebastian, Matias~D Cattaneo, Max~H Farrell, \harvardand\ Rocio
  Titiunik. 2019.
``Regression discontinuity designs using covariates.'' {\em Review of Economics
  and Statistics} 101~(3): 442--451.

\bibitem[Calonico, Cattaneo, \harvardand\ Titiunik(2014)]{calonico2014robust}
Calonico, Sebastian, Matias~D Cattaneo, \harvardand\ Rocio Titiunik. 2014.
``Robust nonparametric confidence intervals for regression-discontinuity
  designs.'' {\em Econometrica} 82~(6): 2295--2326.

\bibitem[Caughey \harvardand\ Sekhon(2011)]{caughey2011elections}
Caughey, Devin, \harvardand\ Jasjeet~S Sekhon. 2011.
``Elections and the regression discontinuity design: Lessons from close US
  house races, 1942--2008.'' {\em Political Analysis} 19~(4): 385--408.

\bibitem[Eggers \harvardand\ Hainmueller(2009)]{eggers2009mps}
Eggers, Andrew~C, \harvardand\ Jens Hainmueller. 2009.
``MPs for sale? Returns to office in postwar British politics.'' {\em American
  Political Science Review} 103~(4): 513--533.

\bibitem[Erikson \harvardand\ Rader(2017)]{erikson2017much}
Erikson, Robert~S, \harvardand\ Kelly Rader. 2017.
``Much ado about nothing: rdd and the incumbency advantage.'' {\em Political
  Analysis} 25~(2): 269--275.

\bibitem[Fisman, Schulz, \harvardand\ Vig(2014)]{fisman2014private}
Fisman, Raymond, Florian Schulz, \harvardand\ Vikrant Vig. 2014.
``The private returns to public office.'' {\em Journal of Political Economy}
  122~(4): 806--862.

\bibitem[Fr{\"o}lich(2007)]{frolich2007regression}
Fr{\"o}lich, Markus. 2007.
``Regression discontinuity design with covariates.'' {\em University of St.
  Gallen, Department of Economics, Discussion Paper}~(2007-32).

\bibitem[Green et~al.(2009)]{green2009testing}
Green, Donald~P, Terence~Y Leong, Holger~L Kern, Alan~S Gerber, \harvardand\
  Christopher~W Larimer. 2009.
``Testing the accuracy of regression discontinuity analysis using experimental
  benchmarks.'' {\em Political Analysis} 17~(4): 400--417.

\bibitem[Hahn, Todd, \harvardand\ Van~der Klaauw(2001)]{hahn2001identification}
Hahn, Jinyong, Petra Todd, \harvardand\ Wilbert Van~der Klaauw. 2001.
``Identification and estimation of treatment effects with a
  regression-discontinuity design.'' {\em Econometrica} 69~(1): 201--209.

\bibitem[Imai(2011)]{imai2011introduction}
Imai, Kosuke. 2011.
``Introduction to the Virtual Issue: Past and Future Research Agenda on Causal
  Inference.'' {\em Political Analysis} 19~(V2): 1--4.

\bibitem[Imbens \harvardand\ Kalyanaraman(2012)]{imbens2012optimal}
Imbens, Guido, \harvardand\ Karthik Kalyanaraman. 2012.
``Optimal bandwidth choice for the regression discontinuity estimator.'' {\em
  The Review of economic studies} 79~(3): 933--959.

\bibitem[James \harvardand\ Stein(1961)]{james1961proc}
James, W, \harvardand\ C Stein. 1961.
``Proc. Fourth Berkeley Symp. Math. Statist. Probab.''  In {\em Estimation with
  quadratic loss}.
Vol.~1. Univ. California Press.

\bibitem[Meier, Van De~Geer, \harvardand\ B{\"u}hlmann(2008)]{meier2008group}
Meier, Lukas, Sara Van De~Geer, \harvardand\ Peter B{\"u}hlmann. 2008.
``The group lasso for logistic regression.'' {\em Journal of the Royal
  Statistical Society: Series B (Statistical Methodology)} 70~(1): 53--71.

\bibitem[Rubin(2005)]{rubin2005causal}
Rubin, Donald~B. 2005.
``Causal inference using potential outcomes: Design, modeling, decisions.''
  {\em Journal of the American Statistical Association} 100~(469): 322--331.

\bibitem[Simon et~al.(2013)]{simon2013sparse}
Simon, Noah, Jerome Friedman, Trevor Hastie, \harvardand\ Robert Tibshirani.
  2013.
``A sparse-group lasso.'' {\em Journal of computational and graphical
  statistics} 22~(2): 231--245.

\bibitem[Skovron \harvardand\ Titiunik(2015)]{skovron2015practical}
Skovron, Christopher, \harvardand\ Roc{\i}o Titiunik. 2015.
``A practical guide to regression discontinuity designs in political science.''
  {\em American Journal of Political Science}: 1--47.

\bibitem[Stein(1956)]{stein1956}
Stein, Charles. 1956.
``Inadmissibility of the Usual Estimator for the Mean of a Multivariate Normal
  Distribution.''  In {\em Proceedings of the Third Berkeley Symposium on
  Mathematical Statistics and Probability, Volume 1: Contributions to the
  Theory of Statistics}, .
Berkeley, Calif. pp.~197--206.
https://projecteuclid.org/euclid.bsmsp/1200501656.

\bibitem[Szakonyi(2018)]{szakonyi2018businesspeople}
Szakonyi, David. 2018.
``Businesspeople in Elected Office: Identifying Private Benefits from
  Firm-Level Returns.'' {\em American Political Science Review} 112~(2):
  322--338.

\bibitem[Tibshirani(1996)]{tibshirani1996regression}
Tibshirani, Robert. 1996.
``Regression shrinkage and selection via the lasso.'' {\em Journal of the Royal
  Statistical Society. Series B (Methodological)}: 267--288.

\bibitem[Tibshirani, Wainwright, \harvardand\
  Hastie(2015)]{tibshirani2015statistical}
Tibshirani, Robert, Martin Wainwright, \harvardand\ Trevor Hastie. 2015.
{\em Statistical learning with sparsity: the lasso and generalizations}.
Chapman and Hall/CRC.

\bibitem[Tihonov(1963)]{tihonov1963solution}
Tihonov, Andrei~Nikolajevits. 1963.
``Solution of incorrectly formulated problems and the regularization method.''
  {\em Soviet Math.} 4: 1035--1038.

\bibitem[Truex(2014)]{truex2014returns}
Truex, Rory. 2014.
``The returns to office in a “rubber stamp” parliament.'' {\em American
  Political Science Review} 108~(2): 235--251.

\bibitem[Wager et~al.(2016)]{wager2016high}
Wager, Stefan, Wenfei Du, Jonathan Taylor, \harvardand\ Robert~J Tibshirani.
  2016.
``High-dimensional regression adjustments in randomized experiments.'' {\em
  Proceedings of the National Academy of Sciences} 113~(45): 12673--12678.

\bibitem[Zou(2006)]{zou2006adaptive}
Zou, Hui. 2006.
``The adaptive lasso and its oracle properties.'' {\em Journal of the American
  statistical association} 101~(476): 1418--1429.

\end{thebibliography}

\end{document}